\documentclass[aps,pra,twocolumn,showpacs,preprintnumbers,amsmath,amssymb,superscriptaddress]{revtex4-1}
\usepackage{graphicx}% Include figure files
\usepackage[latin1]{inputenc}
\usepackage{dcolumn}
\usepackage{bm}

\newcommand{\diff}[0]{\text{d}}

\DeclareMathOperator{\re}{Re}
\DeclareMathOperator{\im}{Im}

\begin{document}

\title{Total internal reflection and evanescent gain}
\author{Jon Olav Grepstad}
\author{Johannes Skaar}\email{johannes.skaar@iet.ntnu.no}
\affiliation{Department of Electronics and Telecommunications, Norwegian University of Science and Technology, NO-7491 Trondheim, Norway}
\affiliation{University Graduate Center, NO-2027 Kjeller, Norway}
%\email{JonOlav.Grepstad@sintef.no}

%\affiliation{Department of Electronics and Telecommunications, Norwegian University of Science and Technology, NO-7491 Trondheim, Norway}
%\affiliation{University Graduate Center, NO-2027 Kjeller, Norway}

\date{\today}% It is always \today, today,

\begin{abstract}
Total internal reflection occurs for large angles of incidence, when light is incident from a high-refractive-index medium onto a low-index medium. We consider the situation where the low-index medium is active. By invoking causality in its most fundamental form, we argue that evanescent gain may or may not appear, depending on the analytic and global properties of the permittivity function. For conventional, weak gain media, we show that there is an absolute instability associated with infinite transverse dimensions. This instability can be ignored or eliminated in certain cases, for which evanescent gain prevails.
\end{abstract}


\maketitle

\section{Introduction}
When light is incident from a high-refractive-index medium onto a low-index medium, it undergoes total internal reflection provided the angle of incidence is larger than a certain critical angle. Total internal reflection is a fundamental physical phenomenon with several famous applications; in particular modern telecommunications rely on optical fibers based on this phenomenon.

Since the tangential electric and magnetic fields must be continuous at the interface, there must be nonzero fields in the low-index medium, even though the incident wave is totally reflected. For lossless/gainless media, these evanescent fields decrease exponentially away from the interface. The presence of evanescent fields in the low-index medium suggests that the reflected wave will sense any perturbation induced there. In particular, if the low-index medium has gain, the reflection response will change compared to the lossless/gainless case. The problem of determining the correct electromagnetic response in the case of an active low-index medium is far from trivial, and has been discussed for 40 years without reaching consensus \cite{romanov72,kolokolov75,callary76,lukosz76,cybulski77,cybulski83,kolokolov98,kolokolov99,fan03,willis08,siegman10}. A key issue is whether the reflectivity may exceed unity (i.e., evanescent gain exists) when the active medium fills the entire half-space. Experiments have indicated that evanescent gain exists \cite{koester66, kogan72, lebedev73, silverman83}. However, it has been argued that the amplified reflection may be due to backreflection from e.g., the boundaries of the active medium \cite{siegman10}.

When the active medium has a finite thickness, it is well known that the overall reflection from the slab may exceed unity. This situation is fairly simple, as there is no need to determine the sign of the longitudinal wavenumber in solving Maxwell's equations for this case; the two waves (with opposite signs) are present simultaneously.

Since there are no gain media with infinite thickness, why examine this case? The answer becomes clear if we formulate a similar question in terms of the refractive index: Why define the refractive index as a separate parameter, when the electromagnetic field in any realistic, bounded structure can be expressed in terms of the permeability and permittivity? 
While the refractive index or longitudinal wavenumber are not needed to obtain the formal solution to Maxwell's equation in a finite slab, it is still useful since it immediately provides information about the involved physics. For example, it predicts whether the medium refracts positively or negatively \cite{nistad08}. Also, assuming darkness for time $t<0$, the solution to Maxwell's equations for a semi-infinite gain medium equals that of a finite slab for times $t$ less than $d/c$, where $d$ is the slab thickness and $c$ is the vacuum velocity of light. Hence, understanding semi-infinite media helps explaining transient phenomena.

\begin{figure}
\begin{center}
\includegraphics[width=5.5cm]{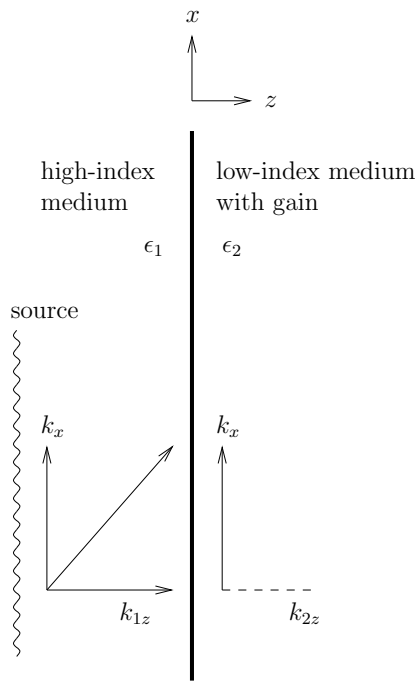}
\end{center}
\caption{A wave is incident from a high-index medium to a low index medium with gain. The source produces a single, spatial frequency $k_x$. The electromagnetic boundary conditions require preservation of the wavenumber $k_x$ parallel to the interface. The longitudinal wavenumbers are denoted $k_{1z}$ and $k_{2z}$. Note that since the excitation is assumed to be causal, it contains a band of frequencies, and therefore also a band of $k_{1z}$'s and $k_{2z}$'s.}
\label{fig:setup}
\end{figure}

\begin{figure}
\begin{center}
\includegraphics[width=7.5cm]{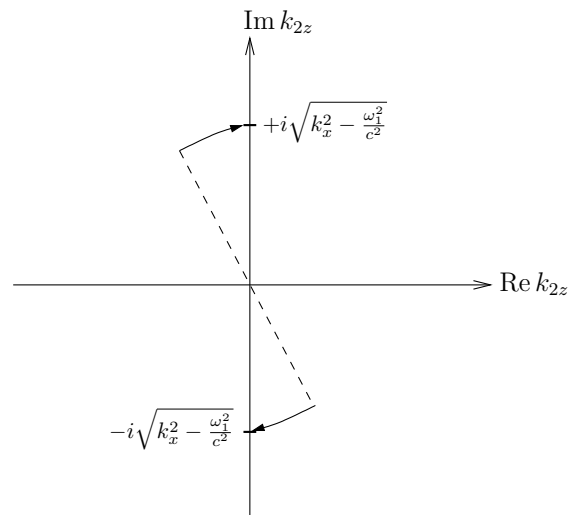}
\end{center}
\caption{The two possible solutions for the wavenumber $k_{2z}$ for monochromatic analysis and a gainy medium. The arrows indicate the two possible wavenumbers in the complex plane, as the gain tends to zero. For a lossy medium, we always have a solution that tends to the upper alternative $+i\sqrt{k_x^2-\omega_1^2/c^2}$ in the limit of zero loss. For simplicity we have taken $\re\epsilon_2=1$ here.}
\label{fig:discont}
\end{figure}

We will now summarize the existing controversy. Assuming well defined frequency-domain fields, Maxwell's equations can be solved in the frequency-domain, using the sign convention $\exp(-i\omega t)$. With respect to Fig. \ref{fig:setup} we define the transverse wavenumber (spatial frequency of the source) $k_x$. For simplicity we assume both media to be nonmagnetic. Let $\epsilon_1$ and $\epsilon_2$ be the relative permittivities of the high-index medium to the left and the low-index medium to the right, respectively. For plane waves, Maxwell's equations require the longitudinal wavenumbers in the high-index and low-index media to be
\begin{subequations}
\begin{align}
k_{1z}&=\pm\sqrt{\epsilon_1\omega^2/c^2-k_x^2}, \label{kzm}\\
k_{2z}&=\pm\sqrt{\epsilon_2\omega^2/c^2-k_x^2}. \label{kz}
\end{align}
\end{subequations}
At some observation frequency $\omega=\omega_1$, we assume $k_x^2<\re\epsilon_1\,\omega_1^2/c^2$ while $k_x^2>\re\epsilon_2\,\omega_1^2/c^2$. Since the high-index medium is passive, we may readily determine the correct sign of the square root in Eq. \eqref{kzm}. For the low-index medium, we assume $\im\epsilon_2<0$ and $|\im\epsilon_2|\ll 1$ (i.e., small gain). The correct sign for the square root in Eq. \eqref{kz} is far from obvious: Either $\im k_{2z}>0$ and $\re k_{2z}<0$, or $\im k_{2z}<0$ and $\re k_{2z}>0$, see Fig. \ref{fig:discont}. None of these solutions are appealing: The first requires the phase velocity and Poynting vector to point {\it towards} the boundary. Since there are no sources at $z=\infty$, one may argue that this scenario cannot be true \cite{siegman10}. The second solution requires that the fields increase exponentially away from the boundary. Also, in the limit of zero gain the fields will increase exponentially as $\exp(z\sqrt{k_x^2-\re\epsilon_2\,\omega_1^2/c^2})$ (see Fig. \ref{fig:discont}), while in the limit of zero loss, the fields decrease exponentially as $\exp(-z\sqrt{k_x^2-\re\epsilon_2\,\omega_1^2/c^2})$. Such a discontinuity seems unphysical \cite{fan03}.

In this work we will first go back to fundamental electromagnetics, to ensure that we use the principle of causality in its most primitive form: No signal can propagate faster than the vacuum velocity of light. After the general analysis in Section \ref{sec:laplace}, we consider conventional, weak gain media in Section \ref{sec:weakgain} and show that they provide evanescent gain. In Section \ref{sec:general} we present an example that demonstrates that not all gain media give evanescent gain; this depends on the medium's global dispersion behavior.

\section{Laplace transform frequency-domain analysis}\label{sec:laplace}
Going back to fundamental electromagnetics, we note that Maxwell's equations, combined with appropriate causal constitutive relations, contain everything necessary to obtain a unique solution. To determine the correct solution, we must be certain that we consider the real, physical situation. The real physical fields are the ones in the time-domain. By requiring the fields to be zero for $t<0$ (see Appendix \ref{sec:instabilities}), we obtain the causal solution to Maxwell's equations. The complex frequency-domain fields are usually found from the time-domain fields by a Fourier transform. However, when there is gain in the system, using the Fourier transform can be perilous, since the field may increase with time. At first sight, any instability seems to be convective in our case. This is however not true: A causal excitation involves an infinite band of frequencies. For a single spatial frequency $k_x$ this means that modes with a wide range of incident angles are involved; in fact even the mode with $k_{2z}=0$ may be excited. This ``side wave'' gets amplified and leads to infinite fields at the boundary. This instability is somewhat artificial, since its existence is dependent on infiniteness in the transverse direction; we will argue below how it can be ignored in certain situations. Nevertheless, within a linear medium framework, Fourier transforms do not necessarily exist. Therefore, as in electronics and control engineering, we generalize the analysis by using the Laplace transform,
\begin{equation}
 E(\omega) = \int_0^\infty \mathcal E(t)\exp(i\omega t)\diff t. \label{laplace}
\end{equation}
In Eq. \eqref{laplace} a sufficiently large value of $\im\omega$ will quench an exponential increase in the time-domain electric field $\mathcal E(t)$, such that the integral converges. (Note that $\omega$ is complex in general, equal to $is$, where $s$ is the conventional Laplace variable.) The inverse transform is given by
\begin{equation}
 \mathcal E(t) = \frac{1}{2\pi}\int_{-\infty+i\gamma}^{+\infty+i\gamma} E(\omega)\exp(-i\omega t)\diff\omega. \label{invlaplace}
\end{equation}
The integral is taken along the line $\omega=i\gamma$, for a sufficiently large, real parameter $\gamma$, above all non-analytic points of $E(\omega)$ in the complex $\omega$-plane. An important observation is the following: The frequency-domain field $E(\omega)$ only has physical meaning through the transforms \eqref{laplace}-\eqref{invlaplace}. Thus, if the field is to be interpreted for all real frequencies, it must be analytic in the upper half-plane $\im\omega>0$. However, as is shown below, if the non-analytic points are located in the upper half-plane, but close to the real axis and far away from the excitation frequency, we can still attribute a physical interpretation to the frequency-domain expressions.

In order to derive the Fresnel equations and determine the sign of $k_{2z}$, it is tempting to start with the response from a slab of finite thickness $d$, and then take the limit $d\to\infty$. For finite $d$ the solution to Maxwell's equations is independent of the sign of $k_{2z}$ in the slab \cite{skaar06,lakhtakia07}. However, for an active slab the multiple reflections may diverge, especially for a large $d$. Thus, for real frequencies, the limit $d\to\infty$ is not necessarily meaningful \cite{skaar06,siegman10}. A way around this, is to evaluate the fields for sufficiently large $\im\omega$, where the frequency-domain fields exist. There, an exponential increase is quenched by the exponential factor $\exp(-\im\omega\, t)$. As a result, we can take the limit $d\to\infty$ \cite{nistad08}. 
%More intuitively, we can expand the field into a geometrical series where each term corresponds to a certain number of round trips in the slab. The first term corresponds to zero round trips. Retaining this term only, we obtain the fields for small $t$, before the far end of the slab has had any effect. As $d\to\infty$, this first term is valid for all times. 
For TE polarization, the Fresnel reflection coefficient $\rho$ and the transmission coefficient $\tau$ (including the propagation factor $\exp(ik_{2z} z)$) become \cite{skaar06,nistad08}
\begin{subequations}
\label{fresneloblique}
\begin{align}
&\rho=\frac{k_{1z}-k_{2z}}{k_{1z}+k_{2z}},\label{fresnelroblique}\\
&\tau=\frac{2k_{1z}}{k_{1z}+k_{2z}}\exp(ik_{2z} z), \label{fresnelsoblique}
\end{align}
\end{subequations}
provided the sign of $k_{2z}$ is determined such that $k_{2z}\to+\omega/c$ as $\im\omega\to\infty$, and $k_{2z}$ is an analytic function of $\omega$. Indeed, even though Eqs. \eqref{fresneloblique} have been derived for large $\im\omega$, we can extend their valid region as follows: The reflected and transmitted frequency-domain fields are given by Eqs. \eqref{fresneloblique} multiplied by the Laplace-transformed incident field. The associated, physical, time-domain fields are obtained by the inverse transform \eqref{invlaplace}. Now, by analytic continuation, we can reduce $\gamma$ until we reach a non-analytic point of Eqs. \eqref{fresneloblique}, without altering $\mathcal E(t)$. If the expressions \eqref{fresneloblique} are analytic in the entire, upper half-plane, we can set $\gamma=0$ and interpret $\rho$ and $\tau$ for real frequencies. On the other hand, if there are non-analytic points in the upper half-plane, the time-domain fields diverge. In that case, real frequencies are not physically meaningful in general.

\section{Weak gain media}\label{sec:weakgain}
To find the actual reflection and transmission response, we first consider conventional weak gain media, with the following assumptions or properties: 
\begin{enumerate}
\item
The permittivity $\epsilon_2(\omega)$ obeys the Kramers--Kronig relations.
\item 
The gain and dispersion is small, so that the permittivity can be written
\begin{equation}
\epsilon_2(\omega) = \bar\epsilon_2 + \Delta\epsilon_2(\omega), \quad|\Delta\epsilon_2(\omega)|\ll \bar\epsilon_2
\end{equation}
for real $\omega$. Here $\bar\epsilon_2$ is required to be a positive constant. In the following we take $\bar\epsilon_2=1$; the analysis can easily be generalized to the case with another $\bar\epsilon_2$. (In the latter case, $\bar\epsilon_2$ is only constant in a wide frequency band including the band where $\Delta\epsilon_2(\omega)$ is nonzero; for very high frequencies it necessarily tends to 1.)
\item
The medium is gainy at the observation frequency $\omega_1$ and the critical frequency $k_xc$. 
\item
Let $\Delta\epsilon_{\max} \equiv \max_{\omega}\left|\Delta\epsilon_2(\omega)\right|$. In a bandwidth $\Delta\epsilon_{\max}k_x c$ around the critical frequency $k_xc$, the permittivity $\epsilon_2(\omega)$ varies slowly:
\begin{equation}
\left|\frac{\diff\epsilon_2}{\diff\omega}\right|<\frac{2}{k_xc} \quad \text{for } |\omega-k_xc|<\frac{\Delta\epsilon_{\max}}{2} k_xc.
\end{equation}
\end{enumerate}
Properties 2 and 4 essentially mean that the gain is weak and the dispersion is small. 

We now solve the equation
\begin{equation}\label{dispersionrel}
\epsilon_2(\omega)\frac{\omega^2}{c^2}=k_x^2,
\end{equation}
to determine whether $k_{2z}$ has branch points in the upper half-plane of the complex $\omega$ plane. Since $\epsilon_2(\omega)$ satisfies the Kramers--Kronig relations, it is analytic in the upper half-plane. The maximum modulus principle of complex analysis \cite{ahlfors} therefore ensures that property 2 is valid also in the upper half-plane, not only at the real frequency axis. Substituting $\epsilon_2(\omega)=1+\Delta\epsilon_2(\omega)$ into Eq. \eqref{dispersionrel} we find 
\begin{equation}\label{disprelsol}
\omega=\pm k_xc\left(1-\frac{\Delta\epsilon_2(\omega)}{2}\right) 
\end{equation}
in the upper half-plane, since $|\Delta\epsilon_2(\omega)|\ll 1$. Thus, every solution to the dispersion relation in the upper half-plane is located within a distance $(\Delta\epsilon_{\max}/2)k_xc$ from the critical frequencies $\pm k_xc$.

We therefore examine the region around $k_xc$ in more detail. If there were two solutions $\omega_a$ and $\omega_b$ to the dispersion relation, then Eq. \eqref{disprelsol} would predict that 
\begin{equation}
\Delta\epsilon_2(\omega_a)-\Delta\epsilon_2(\omega_b)=\frac{2}{k_xc}(\omega_b-\omega_a).
\end{equation}
By property 4 this is impossible unless $\omega_b=\omega_a$. Thus there is a unique solution to Eq. \eqref{dispersionrel} in the first quadrant, located in the vicinity of $k_xc$:
\begin{equation}\label{finsol}
\omega=k_xc+\delta'+i\delta, 
\end{equation}
where 
\begin{equation}
\delta'=-\frac{\re\Delta\epsilon_2(\omega)}{2}k_xc \ \text{ and }
\delta=-\frac{\im\Delta\epsilon_2(\omega)}{2}k_xc.
\end{equation}
In addition there is a mirrored solution in the second quadrant, located at $\omega=-k_xc-\delta'+i\delta$. Note that
\begin{equation}\label{deltamax}
\delta\leq\frac{\Delta\epsilon_{\max}}{2}k_xc.
\end{equation}

\begin{figure}
\begin{center}
\includegraphics[width=7.5cm]{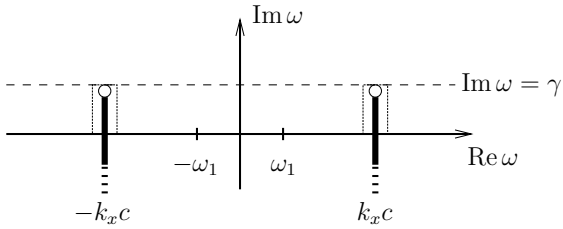}
\end{center}
\caption{The complex $\omega$-plane. For conventional, weak gain media, there are branch points right above $\omega=\pm (k_xc+\delta')\approx\pm k_xc$. The branch cuts can be chosen arbitrarily; however, the shown, vertical cuts minimize the integral around the part of the branch cuts in the upper half-plane.} 
\label{fig:omega}
\end{figure}

In the expression for $k_{2z}$ and the Fresnel coefficients \eqref{fresneloblique}, these solutions appear as branch points. Hence, when evaluating the physical time-domain fields by the inverse Laplace transform, we must integrate above the associated branch cuts, from $-\infty+i\gamma$ to $+\infty+i\gamma$, see Fig. \ref{fig:omega}. By path deformation this path is the same as the path from $-\infty$ to $\infty$ plus the paths around the branch cuts in the upper half-plane (Fig. \ref{fig:omega}). Thus we may use the inverse Fourier transform to determine the time-domain fields, but only if we add the integrals around the branch cuts. Due to the exponential factor $\exp(-i\omega t)$, the integrals around the branch cuts diverge and dominate after some time.

The divergence of the time-domain fields can be explained as follows. Any causal excitation involves an infinite frequency band. For example, the Laplace transform of a unit-step-function modulated cosine, $u(t)\cos(\omega_1 t)$, is $i\omega/(\omega^2-\omega_1^2)$. Thus, it is nonzero for all finite $\omega\neq 0$. One of these frequencies is the branch-point frequency for which $k_{2z}=0$, that is, $\omega\approx k_xc+i\delta$. This frequency is complex; the imaginary part $\delta$ means that the associated eigenmode is a growing wave with envelope $\exp(\delta t)$. Physically, a wave with $k_{2z}=0$ propagates along the boundary. Because the medium is gainy, this side wave picks up gain on its way. Consider a fixed observation point, e.g. the point $z=0^+$ and $x=0$. Since the medium and the excitation are unbounded in the transverse $x$-direction, there are side waves that start arbitrarily far away from the observation point. Thus the field at the observation point diverges. As the field in medium 2 becomes infinite, the field in medium 1 is infinite as well. Since the field at a fixed point in space diverges and the instability is not a result of amplified, multiple reflections, the instability for the system in Fig. \ref{fig:setup} can be classified as an absolute instability \cite{sturrock,briggs,skaar06}.

This instability could be eliminated (or converted into a convective instability) by limiting the extent of the gain medium in the transverse direction with an absorbing boundary. Alternatively, the incident wave itself could be limited in the $x$-direction, leading to an infinite spectrum of $k_x$ modes (see Appendix A and Ref. \cite{kolokolov99}). Rather than imposing such remedies, we will simply calculate the time-domain fields by an inverse Laplace transform above the branch cuts. If the excitation frequency $\omega_1$ is sufficiently remote from the branch points, the side wave with $k_{2z}=0$ is only excited very weakly, and can be neglected up to a certain time. The condition that the excitation frequency is remote from $k_x c$ means that the incident angle is not close to the critical angle. This condition is imperative in order to distinguish between the reflected wave, with an angle of reflection equal to the angle of incidence, and the wave associated with the growing side wave, with ``reflection'' (or propagation) angle equal to the critical angle. 

The reflected time-domain field for the excitation
%\footnote{A physical excitation is real and can be represented by, e.g., $u(t)\exp(-i\omega_1 t)+u(t)\exp(i\omega_1^* t)$. It is convenient to consider the positive frequency excitation separately.} 
$u(t)\exp(ik_x x-i\omega_1 t)$, with Laplace transform $\exp(ik_x x)/(i\omega_1-i\omega)$, is given by
\begin{equation}
 \mathcal E_{\rho}(x,t) = \frac{1}{2\pi}\int_{-\infty+i\gamma}^{+\infty+i\gamma} \frac{k_{1z}-k_{2z}}{k_{1z}+k_{2z}} \frac{\exp(ik_x x-i\omega t)}{i\omega_1-i\omega}\diff\omega, \label{reflfield}
\end{equation}
at $z=0$. 
The integral \eqref{reflfield} can be evaluated by a generalized version of the residue theorem, in which we find the contour integral around all poles and branch cuts of the integrand in half-plane $\im\omega<\gamma$. Provided $\omega_1$ is sufficiently remote from any resonances of the two media, the transients due to all poles and branch cuts for $\im\omega<0$ can be ignored. Alternatively, for times larger than the maximum inverse bandwidth $\Gamma^{-1}$ of the resonances, the transients will have died out. Then the reflected field for $x=0$ is given by
\begin{equation}\label{freqdomainsol}
\mathcal E_{\rho}(0,t)=\frac{k_{1z}-k_{2z}}{k_{1z}+k_{2z}}\exp(-i\omega_1 t) + \mathcal E_\text{bc}(0,t), \quad t\gtrsim \Gamma^{-1},
\end{equation}
where the wavenumbers $k_{1z}$ and $k_{2z}$ have been evaluated at the frequency $\omega_1$. The term $\mathcal E_{\text{bc}}(0,t)$ is the integral \eqref{reflfield} around the two branch cuts above $\omega=\pm k_xc$. This integral is bounded by
\begin{equation}\label{bcbound}
\left| \mathcal E_{\text{bc}}(0,t) \right| \leq \frac{\text{const}}{k_xc-\omega_1}\cdot\exp(\delta t).
\end{equation}
Here, the constant depends on the specifics of the active medium (see Appendix C). In other words, for $\Gamma^{-1}\lesssim t \lesssim \delta^{-1}$ and provided $\omega_1$ is not too close to $k_xc$, we can ignore $\mathcal E_{\text{bc}}(0,t)$. Then the reflected field is well described by the first term in Eq. \eqref{freqdomainsol}.

\begin{figure}
\begin{center}
\includegraphics[width=6.5cm]{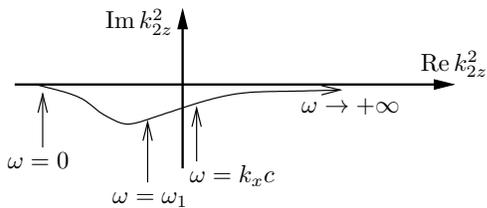}
\end{center}
\caption{The function $k_{2z}^2(\omega)=\epsilon_2(\omega)\omega^2/c^2-k_x^2$ for a typical gain medium, plotted in the complex $k_{2z}^2$-plane. To identify $k_{2z}$, we require it to be $+\omega/c$ at $\omega=\infty$, continuous as $\omega$ decreases towards zero, except at the branch cut at $\omega = k_x c$ where it changes sign.}
\label{fig:k2z2}
\end{figure}

We can now answer the question about the existence of evanescent gain. To obtain Eq. \eqref{freqdomainsol}, we have only considered two branch cuts in the upper half-plane; these are the necessary branch cuts due to the zeros of $\epsilon_2(\omega)\omega^2/c^2 - k_x^2$. We must ensure that the integrand in Eq. \eqref{reflfield} is analytic everywhere else in the upper half-plane. That is, the sign of $k_{2z}$ must be determined such that $k_{2z}$ is analytic everywhere, except at the two branch cuts in the upper half-plane. Since $k_{2z}\to +\omega/c$ as $\omega\to+\infty$, we can determine the sign by decreasing $\omega$ from $+\infty$ to $\omega_1$, ensuring that $k_{2z}$ is continuous everywhere except at $\omega=k_xc$ where it changes sign. From Fig. \ref{fig:k2z2} we find that $\im k_{2z}>0$ at the observation frequency $\omega_1$. Hence, for weak conventional gain media, provided the ``reflected'' field from the side wave can be ignored, evanescent gain is possible. This result is consistent with \cite{kolokolov98,kolokolov99}, and the time-domain simulations in \cite{willis08} where the dispersion of the medium is discarded.

\begin{figure}
\begin{center}
\includegraphics[width=6cm]{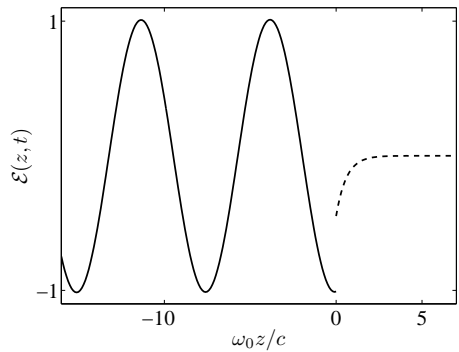}
\end{center}
\caption{The reflected electric field (solid line, $z<0$) and transmitted electric field (dashed line, $z>0$) for a plane wave incident to a weak Lorentzian gain medium \eqref{invlorentz}. The parameters used: $F=0.01$, $\Gamma=0.1\omega_0$, $k_xc=2\omega_0$, $\omega_1=\omega_0$, $\gamma=0.001$, and $\epsilon_1=4.7$. The field is plotted for $x=0$ and $\omega_0t=10^3$, and normalized to the incident field. The amplitude of the reflected field is 1.01. Note that the field is discontinuous at $z=0$ because the incident wave is not included.}
\label{fig:ev}
\end{figure}
In Fig. \ref{fig:ev} we plot the reflected and transmitted electric field for a weak Lorentzian medium, after the transients have died out, and before the side wave dominates. The reflected field was computed by Eq. \eqref{reflfield}, including the propagation factor $\exp(-ik_{1z}z)$ in the integral. The transmitted field was computed with the same equation, but with $\tau$ instead of $\rho\exp(-ik_{1z}z)$ in the integral (see Eq. \eqref{fresneloblique}). For $z>0$ we clearly see an evanescent decaying field, while the reflected field for $z<0$ is larger than unity.

It is interesting to examine the situation when we approach the critical angle associated with the frequency $\omega_1$. If we insist on using only the first term of Eq. \eqref{freqdomainsol} in this case, a simple calculation shows that the power reflectance would have been bounded by $(\sqrt 2+1)^2\approx 5.83$ at the critical angle. Also, the wavenumber $k_{2z}$ and the reflected field would be discontinuous as we pass the critical angle. This is clearly a paradox, as the branch cuts were chosen arbitrarily. The dilemma is resolved by noting that the entire Eq. \eqref{freqdomainsol} must be used in this domain; both terms naturally coexist and cannot be separated. As we approach the critical angle, $\mathcal E_\text{bc}(0,t)$ becomes comparable to or larger than the first term in Eq. \eqref{freqdomainsol}, for all times. A different choice of branch cuts will alter each of these contributions, but the sum remains the same. For finite transverse dimension, the side wave's contribution to the ``reflected'' field does not necessarily diverge any more; however, the intensity of the reflected field can be arbitrarily large as the dimension is increased, or if the reflections from the transverse end facets are large.

\section{General gain media}\label{sec:general}
More sophisticated gain media can be constructed, at least in principle, that behave differently compared to the conventional weak gain media. We will here show that we can obtain a near-imaginary $k_{2z}$ with negative imaginary part at an observation frequency $\omega_1<k_x c$. Consider the permittivity
\begin{equation}\label{epsilon2eng}
 \epsilon_2(\omega) = \frac{(\omega-N)(\omega+N^*)}{(\omega-P)(\omega+P^*)}+\frac{\omega_2^2}{\omega^2},
\end{equation}
where the complex numbers $N$ and $P$ are located in the lower half-plane, and $\omega_2$ is a real constant. The longitudinal wavenumber satisfies $k_{2z}^2(\omega) = \epsilon_2(\omega)\omega^2/c^2-k_x^2$, which gives
\begin{equation}\label{k2z2eng}
 k_{2z}^2(\omega) = \frac{\omega^2(\omega-N)(\omega+N^*)}{c^2(\omega-P)(\omega+P^*)}+\frac{\omega_2^2}{c^2}-k_x^2.
\end{equation}
Choosing $\omega_2=k_xc$, we can tailor the frequency dependence of $k_{2z}^2$ by carefully selecting the locations of zeros and poles. Let $N = n - iC$ and $P = p - iC$, where $C>0$. All poles and zeros are now located in the (closed) lower half-plane. For $\omega>0$, assuming $C \ll n,p$, the longitudinal wavenumber can be written 
\begin{equation} \label{k2z2engA}
  k_{2z}^2(\omega) = A(\omega) \left( B(\omega)+iC(n-p) \right)
\end{equation}
for real functions $A(\omega)$ and $B(\omega)$; in addition $A(\omega)>0$. Hence, for $n<p$, $\im k_{2z}^2<0$ for all positive frequencies. Since $k_{2z}$ is analytic in the upper half-plane of $\omega$, and since $k_{2z}\to+\omega/c$ as $\omega\to\infty$, $k_{2z}$ will be located in the forth quadrant of the complex $k_{2z}$-plane, i.e., $\re k_{2z}>0$ and $\im k_{2z}<0$ for all $\omega > 0$. A proper evanescent or ``anti-evanescent'' wave has $\left|\text{Re}(k_{2z})/ \text{Im}(k_{2z})\right| \ll 1$, so we search for values of $\omega_1$ satisfying this requirement. 
\begin{figure}[!t]
\centering \includegraphics[width=8.08cm]{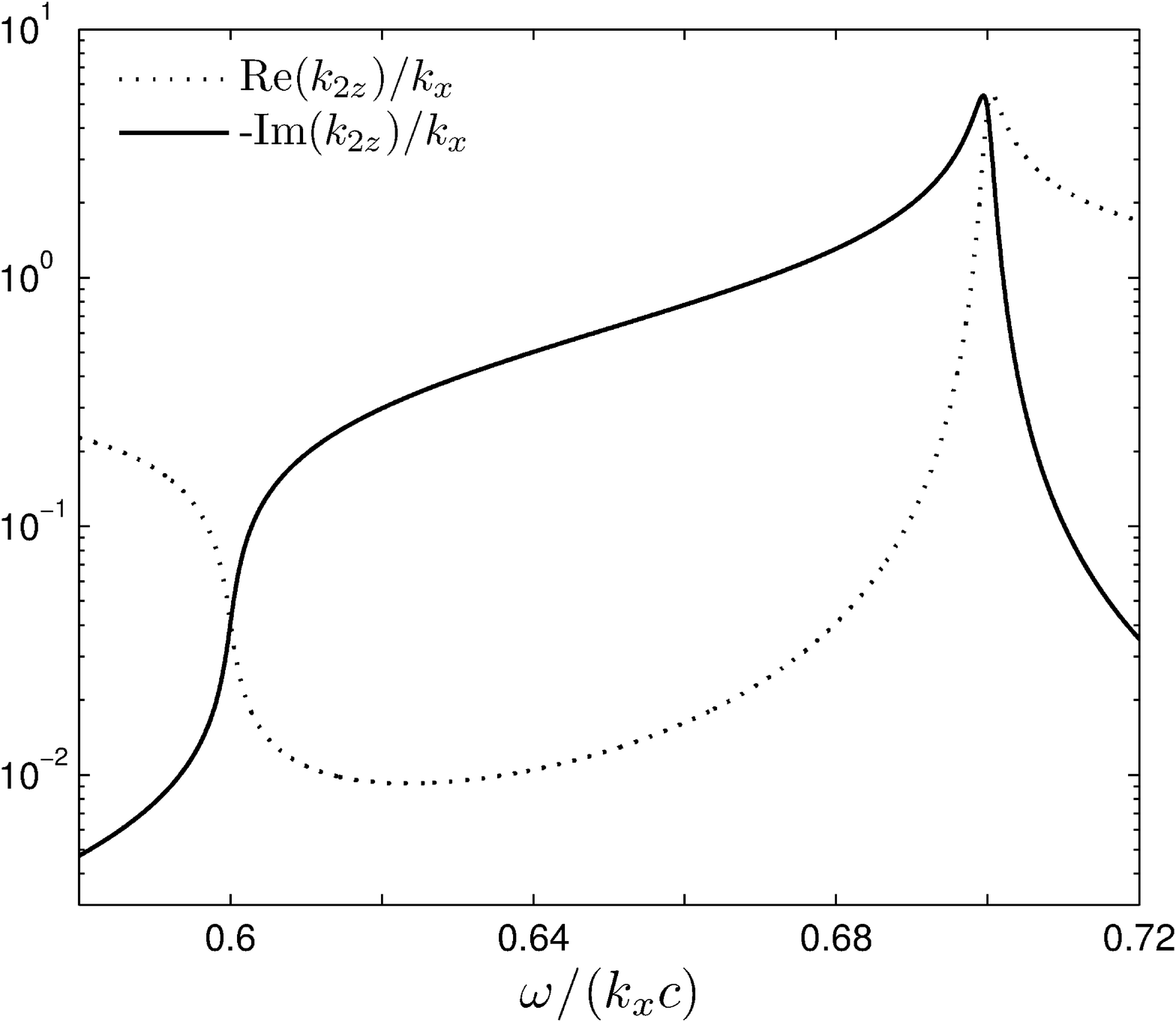}
\caption{The real and imaginary parts of $k_{2z}(\omega)=\sqrt{\epsilon_2\omega^2/c^2-k_x^2}$, with $\epsilon_2$ given by Eq. \eqref{epsilon2eng}. We have set $\omega_2 = k_x c$, $N=k_x c(6/10-i/1000)$ and $P=k_x c(7/10-i/1000)$.}
\label{fig:k2z4q}
\end{figure}
Analyzing Fig. \ref{fig:k2z4q}, there exists an $\omega_1$ where $\left|\text{Re}(k_{2z})/ \text{Im}(k_{2z})\right| \ll 1$ for $n < \omega_1 < p$. We have hence found a medium for which $k_{2z}$ describes an ``anti-evanescent'' wave in a finite frequency range. 

Any realistic incident wave contains a spectrum of wavenumbers $k_x$. While there are no zeros of $k_{2z}^2$ in the upper half-plane for the particular $k_x$ considered above, this is not the case for all possible $k_x$. Thus, also for this medium there are growing waves. The fact that the medium has large gain, and the presence of instabilities, mean that it is very challenging to observe the ``anti-evanescent'' response in practice. In principle, however, up to a certain time the amplitude of the instabilities can be limited by ensuring a narrowbanded spectrum of incident $k_x$'s. Formally, if $\sigma$ is the width of the incident wave, and $\mathcal E_\sigma(x,z,t)$ is the resulting electric field, $\lim_{\sigma\to\infty}\mathcal E_\sigma(x,z,t)$ tends to the ``anti-evanescent'' response as $t\to\infty$, while $\lim_{t\to\infty}\mathcal E_\sigma(x,z,t)=\infty$ for any finite $\sigma$.

The permittivity \eqref{epsilon2eng} has a double pole at $\omega=0$. While the medium is causal in principle, the medium might be easier to realize if the pole is moved slightly away from the origin, into the lower half-plane. It turns out that this modification does not alter the permittivity function signifiantly, in the frequency range of interest. Also, if desired, the behavior at $\omega=\infty$ can be adjusted along the lines described in Ref. \cite{skaar06}.

\begin{figure}
\centering \includegraphics[width=6cm]{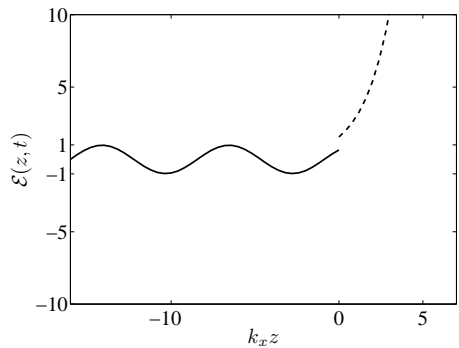}
\caption{The reflected electric field (solid line, $z<0$) and transmitted electric field (dashed line, $z>0$) for a plane wave incident to the gain medium described by Eq. \eqref{epsilon2eng} and Fig. \ref{fig:k2z4q}. The field is plotted for $x=0$ and $k_xc t=10^5$, and normalized to the incident field. The amplitude of the reflected field is 0.98. The parameters used: $\omega_1=0.65 k_x c$, $\gamma=10^{-6}$, and $\epsilon_1=4$.}
\label{fig:antiev}
\end{figure}

In Fig. \ref{fig:antiev} we plot the reflected and transmitted field for the gain medium \eqref{epsilon2eng}, calculated with the inverse Laplace transform for a sufficiently large $t$ when the transients have died out. Only a single $k_x$ has been excited. The reflection amplitude is 0.98, and the transmitted field is an exponentially increasing function of $z$. While realizable in principle, the example is highly unrealistic: To observe a behavior similar to that in Fig. \ref{fig:antiev}, $t$ must be at least of the order of $10^2 (k_xc)^{-1}$; otherwise the transients would disturb the picture. Any realistic gain medium has finite thickness. However, to act as a semi-infinite medium, the thickness $d$ of the gain medium must satisfy $d>ct$, or $k_x d \gtrsim 10^2$, such that the light has not reached the back end. With the ``anti-evanescent'' growth rate in Fig. \ref{fig:antiev}, this would imply unphysically large fields (or in practice, nonlinear gain saturation). Hence, if the ``anti-evanescent'' behavior is to be observed experimentally, one would need to construct a medium where the transients die out rapidly, and/or a medium which leads to a sufficiently small $|\im k_{2z}|$. At the same time the medium must violate the conditions in Sec. \ref{sec:weakgain}; that is, it must have large gain and/or large dispersion for some frequencies.

\section{Conclusion}
We have considered the case where light is incident from a high-index medium to a low-index medium with gain, generalizing the situation with total internal reflection.

In principle, it is apparent that both solutions ($k_{2z}$ in the second and fourth quadrant of the complex plane) can be attained with a suitably engineered medium. In other words, evanescent gain may or may not be the case, dependent on the detailed permittivity function. This demonstrates the fact that the sign of $k_{2z}$ cannot be determined from the electromagnetic parameters at a single frequency, but must be identified from the entire frequency domain dependence, after a check of possible non-analytic points (instabilities) in the upper half-plane of complex frequency.

For conventional, weak gain media, we have seen that there is an absolute instability associated with infinite transverse dimensions. In some cases this instability can be eliminated or ignored; then evanescent gain prevails.

\appendix
\section{Finite incident beam and finite size medium}
To origin from a realistic source, an incident beam should not only be causal, but also be of finite width. We will here describe how to model an incident beam using standard Fourier optics, and argue that even for active media, we are allowed to interchange the order of integration with respect to transverse wavenumber $k_x$ and frequency $\omega$. Thus we can treat a causal excitation of each $k_x$ separately.

Let $\mathcal E(x,t)$ be the incident TE field at the interface between the high-index medium and the active low index medium. Performing a Laplace transform $t\to\omega$ followed by a Fourier transform $x\to k_x$, we obtain the transformed field $E(k_x,\omega)$. The inverse transform is given by
\begin{align}
 &\mathcal E(x,t) 
  =\frac{1}{(2\pi)^2}\int_{-\infty+i\gamma}^{\infty+i\gamma}\diff\omega e^{-i\omega t} \int_{-\infty}^{\infty}\diff k_x E(k_x,\omega)e^{ik_xx} \nonumber\\
  &=\frac{1}{(2\pi)^2}\int_{-\infty}^{\infty}\int_{-\infty}^{\infty} E(k_x,\omega'+i\gamma)e^{-i\omega' t+\gamma t}e^{ik_xx}\diff k_x\diff \omega'.
  \label{invLaplFour}
\end{align}

By Fubini's theorem we may interchange the order of integration in Eq. \eqref{invLaplFour}, provided $E(k_x,\omega'+i\gamma)$ is absolute integrable with respect to $k_x$ and $\omega'$. This is the case assuming that the incident field is sufficiently smooth with respect to $t$ and $x$. For example, taking the incident wave to be $a(x)e^{iK_xx}b(t)e^{-i\omega_1 t}$, the transformed field becomes $E(k_x,\omega'+i\gamma)=A(k_x-K_x)B(\omega'-\omega_1+i\gamma)$, where $A$ is the Fourier transform of $a$, and $B$ is the Laplace transform of $b$. Here we assume that $b(t)=0$ for $t<0$. If $a$ and $b$ are continuous, $A$ and $B$ are absolute integrable.

We can repeat the above argument for the total field (incident + reflected, and transmitted). Assuming no superexponential instabilities, the total field is uniformly bounded: 
\begin{equation}
|\mathcal E(x,z,t)|\leq C\exp(\gamma t), 
\end{equation}
for positive constants $C$ and $\gamma$. Then the transforms $t\to\omega$ followed by $x\to k_x$ exist, and we can express the total field in the form \eqref{invLaplFour}. The total field is determined using the wave equation. In order to consider each mode $k_x$ separately, we interchange the order of integration for each term in the wave equation. To do so, we require the second order derivatives with respect to $t$ and $x$ to be continuous. 

It remains to prove that our solution is consistent with this requirement. From the theory in Sec. \ref{sec:laplace}, we find the solution for each $k_x$, given a sufficiently smooth incident field. For this solution, the Fresnel equations show that the reflection and transmission coefficients tend to zero and unity, respectively, as $|\omega'|\to\infty$ or $|k_x|\to\infty$. Therefore the reflected and transmitted field in the $(\omega,k_x)$-domain adopt any absolute integrability property from the incident field.

In our analysis the incident field $u(t)\exp(ik_xx-i\omega_1t)$ is not continuous. Hence, strictly speaking, the above described method cannot be used. However, by smoothing the discontinuity around $t=0$, we can make  the field and its second order derivative continuous. This modification will not affect the discussion in general, since a slower transient will reduce the bandwidth. Thus the side waves are excited weaker, such that inequality \eqref{bcbound} is satisfied with an even larger margin.

\begin{figure}[!t]
\centering \includegraphics[width=4.5cm]{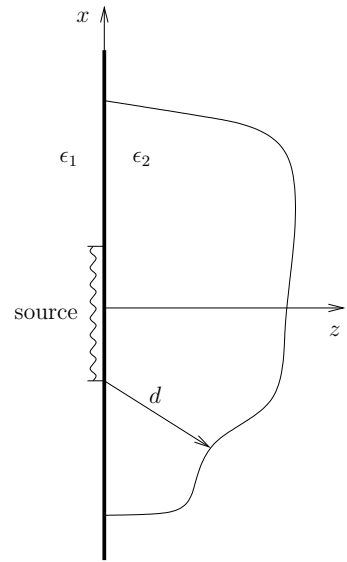}
\caption{\small The semi-infinite gain medium can be replaced by a finite size gain medium, provided we only consider times $t<d/c$.} 
\label{fig:finite}
\end{figure}
In a real experiment, not only the beam width, but also the size of the active medium itself, must be finite. Provided the fields never reach the end of the structure in the time window of interest, the fields will be identical to those in a semi-infinite active medium. Thus we can consider a setup as in Fig. \ref{fig:finite}, where the least distance from the incident beam to the boundary is $d$. For $t<d/c$ the fields will be the same as if the finite-size medium were replaced by a semi-infinite medium.

\section{Instabilities in infinite media}\label{sec:instabilities}
It is convenient to divide instabilities into two categories, convective and absolute instabilities (see e.g. \cite{sturrock,briggs}). Media with absolute instabilities are often regarded as impractical for small-signal, linear applications, since for an unbounded medium the fields diverge even at a fixed point in space. In contrast, media with convective instabilities are useful in the linear regime. Here the fields do not diverge at a fixed point in space; the growing wave is rather convected away.

However, even in the case with only convective instabilities, there may be fundamental problems in the case where the medium occupies an infinite region or half-space: Any small perturbation may propagate an infinite distance, thus picking up an infinite amount of gain. In our analysis we assume that the active medium is dark for $t<0$. It is not clear whether this is possible, not even in principle, since perturbations in the remote past would not die out but rather increase exponentially.

The remedy is motivated by practical considerations. In an experiment, the active medium must have finite size in all directions. For a medium without absolute instabilities and with a given maximum size $d$, there will be no instabilities provided the gain is sufficiently weak. Examples of such configurations include optical amplifiers, and laser resonators with pumping below threshold. When there are no instabilities, we can turn on the pump in remote past such that the perturbations have died out before $t=0$. For $0<t<d/c$ we can still regard the medium as semi-infinite, since, as seen from Fig. \ref{fig:finite}, it makes no difference.

\section{Determining the reflected time-domain field}
\begin{figure*}%[!t]
\centering \includegraphics[width=11cm]{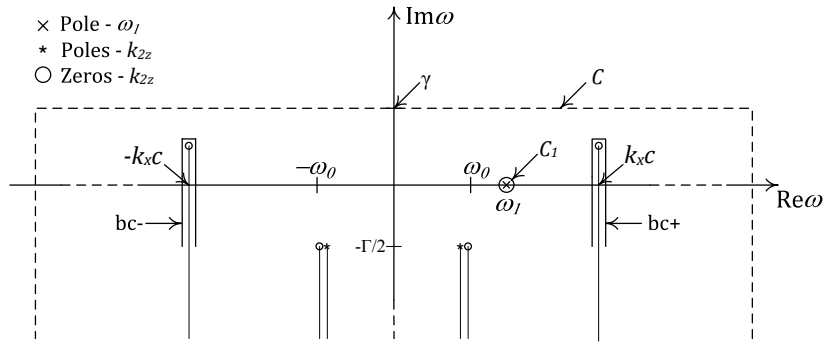}
\caption{\small Circles and stars mark zeros and poles of $k_{2z}$ respectively. The cross marks the pole at $\omega=\omega_1$. Branch cuts are arbitrarily chosen to lie parallel to the imaginary axis, extending into the lower half-plane $\text{Im}(\omega) < 0$. The integration path $C$ is shown with the dashed line. Contributing branch cuts and poles are enclosed by paths $C_1$, bc$-$ and bc$+$. For $t \gtrsim 2/\Gamma$, we have $\oint_{C} f(\omega)\diff\omega \approx \oint_{C_1} f(\omega)\diff\omega + \int_{\text{bc}-} f(\omega)\diff\omega + \int_{\text{bc}+} f(\omega)\diff\omega$.} 
\label{fig:intpath2}
\end{figure*}
Here we will calculate the reflected field in the time-domain, when the gain medium is described by a weak, inverted Lorentzian function:
\begin{equation}\label{invlorentz}
 \epsilon_2(\omega)=1-\frac{F\omega_0^2}{\omega_0^2-\omega^2-i\omega\Gamma}.
\end{equation}
In Eq. \eqref{invlorentz} $F$, $\omega_0$, and $\Gamma$ are positive parameters, describing the resonance strength, frequency, and bandwidth, respectively. The physical, time-domain reflected field at $z = x = 0$ is given by the inverse Laplace transform \eqref{reflfield}, repeated for convenience here:
\begin{equation} \label{reflfieldA}
 \mathcal E_{\rho}(0,t) = \frac{1}{2\pi}\int_{-\infty+i\gamma}^{+\infty+i\gamma} \frac{k_{1z}-k_{2z}}{k_{1z}+k_{2z}} \frac{\exp(-i\omega t)}{i\omega_1-i\omega}\diff\omega.
\end{equation}
The field can be interpreted by evaluating integral \eqref{reflfieldA} by a generalized version of the residue theorem. We here recognize that integrating along path $-\infty +i\gamma$ to $+\infty +i\gamma$, is the same as integrating around all branch cuts and poles. The denominator $k_{1z}+k_{2z}$ does not have any zeros, provided the permittivity $\epsilon_1$ can be considered constant and larger than unity in the frequency range of interest. Thus we only need to consider the branch cuts extending from branch points of $k_{1z}$ and $k_{2z}$, and the pole at $\omega=\omega_1$. Note that the branch cuts are arbitrary, as long as they extend from the branch points. We let all branch cuts lie parallel to the imaginary axis, towards $\im\omega=-\infty$. See illustration in Fig. \ref{fig:intpath2}. The branch points of $k_{1z}$ are located far away from (and below) the real frequency axis, provided the medium's bandwidth is sufficiently large. The wavenumber $k_{2z}$ has two branch points in the upper half-plane, located immediately above $\omega=\pm k_x c$. In addition there are four branch points located below the real frequency axis, with imaginary parts $-\Gamma/2$; two simple zeros and two simple poles. The integrals around the latter four branch cuts decay with time constant at most $2/\Gamma$. Thus, for $t \gtrsim 2/\Gamma$, the only contributing terms are the residue of the pole at $\omega_1$, and the contribution $\mathcal E_\text{bc}(0,t)$ from the two remaining branch cuts of $k_{2z}$:
\begin{equation} \label{freqdomainsolA}
\mathcal E_{\rho}(0,t)=\frac{k_{1z}-k_{2z}}{k_{1z}+k_{2z}}\exp(-i\omega_1 t) + \mathcal E_\text{bc}(0,t).
\end{equation}
Here $k_{1z}$ and $k_{2z}$ have been evaluated at the frequency $\omega_1$. We write $\mathcal E_{\text{bc}}(0,t)=\mathcal E_{\text{bc}-}(0,t)+\mathcal E_\text{bc+}(0,t)$, where $\mathcal E_{\text{bc}-}(0,t)$ and $\mathcal E_\text{bc+}(0,t)$ are the contributions from the branch cuts in the left and right half-planes, respectively.

Assuming $F \ll 1$, $\Gamma \ll \omega_0$ and $\sqrt{2} \omega_0<k_x c$, the branch cut in the right half-plane extends from approximately $\omega = k_xc+i \delta$ to $\omega = k_xc-i \infty$, where $\delta \leq F\Gamma$. Then, for $t \gtrsim 2/\Gamma$ 
\begin{align} \label{branchcut}
 	\mathcal E_\text{bc+}(0,t) & \approx \frac{1}{2\pi}\int_{k_xc-i\frac{\Gamma}{2}}^{k_xc+i\delta} \rho_\text{l}(\omega) \frac{\exp(-i\omega t)}{i\omega_1-i\omega} \diff\omega \nonumber\\
	&- \frac{1}{2\pi}\int_{k_xc-i\frac{\Gamma}{2}}^{k_xc+i\delta} \rho_\text{r}(\omega) \frac{\exp(-i\omega t)}{i\omega_1-i\omega} \diff\omega. 
\end{align}
Here subscripts l and r indicate that $\rho(\omega)$ is discontinuous when crossing the branch cut, denoting the left and right side of the branch cut respectively. We further define $f_\text{l,r}(\omega)=k_{2z}/k_{1z}$. Since $k_{2z}$ is small in the vicinity of $k_x c$, by first order approximation $\rho_\text{l,r}(\omega) = 1-2f_\text{l,r}(\omega)$, where $f_\text{r}(\omega)= -f_\text{l}(\omega)$. The integral \eqref{branchcut} can now be simplified:
\begin{equation} \label{branchcutb}
 \mathcal E_\text{bc+}(0,t) \approx \frac{-2}{i\pi}\int_{k_xc-i\frac{\Gamma}{2}}^{k_xc+i\delta} f_\text{l}(\omega) \frac{\exp(-i\omega t)}{\omega_1-\omega}\diff\omega, 
\end{equation}
In order to obtain a manageable expression for $f_\text{l}(\omega)$, it is useful to express $k_{2z}^2$ as a function of its zeros and poles. With poles denoted by subscript $p$, and zeros denoted by subscripts $k_x c$ and $\omega_0$ (indicating the location along the real frequency axis), $k_{2z}^2$ appears as
\begin{equation} \label{k2zbypole}
  k_{2z}^2 = \frac{(\omega-\omega_{\omega_0})(\omega+\omega_{\omega_0}^*)(\omega-\omega_{k_x c})(\omega+\omega_{k_x c}^*)}{c^2(\omega-\omega_p)(\omega+\omega_p^*)}.
\end{equation}
Identifying $\delta = \text{Im}(\omega_{k_x c})$ and $\omega_i = \text{Im}(\omega)$, and recognizing that $(\omega - \omega_{\omega_0})/(\omega-\omega_p) \approx 1$ at $\omega = \omega_{k_x c}$, Eq. \eqref{k2zbypole} can be simplified: $k_{2z}^2 \approx -i2k_x(\delta-\omega_i)/c$. This gives 
\begin{equation}
|f_\text{l}(k_xc+i\omega_i)| \approx \sqrt{2(\delta-\omega_i)/(k_x c(\epsilon_1-1))}. 
\end{equation}
For $k_xc-\omega_1\gg\Gamma$, we can now find an upper bound of integral \eqref{branchcutb} by noting that $\sqrt{\delta - \omega_i} \leq \sqrt{F \Gamma + \Gamma/2}$ for all $\omega_i$ considered. We can estimate $\mathcal E_{\text{bc}-}(0,t)$ similarly, yielding the bound
\begin{equation} \label{branchcutd}
  \left| \mathcal E_{\text{bc}}(0,t) \right| \leq \frac{\Gamma^{3/2}}{(k_x c-\omega_1)\sqrt{k_x c(\epsilon_1-1)}} \left( e^{F \Gamma t}-e^{-\Gamma t/2} \right) 
\end{equation}
Consequently for $2/\Gamma \lesssim t \lesssim 1/F\Gamma$, the field is well described by the first term in Eq. \eqref{freqdomainsolA}.

%\bibliography{nbib}% Produces the bibliography via BibTeX.

\begin{thebibliography}{10}%
\makeatletter
\providecommand \@ifxundefined [1]{%
 \ifx #1\undefined \expandafter \@firstoftwo
 \else \expandafter \@secondoftwo
\fi
}%
\providecommand \@ifnum [1]{%
 \ifnum #1\expandafter \@firstoftwo
 \else \expandafter \@secondoftwo
\fi
}%
\providecommand \enquote [1]{``#1''}%
\providecommand \bibnamefont  [1]{#1}%
\providecommand \bibfnamefont [1]{#1}%
\providecommand \citenamefont [1]{#1}%
\providecommand\href[0]{\@sanitize\@href}%
\providecommand\@href[1]{\endgroup\@@startlink{#1}\endgroup\@@href}%
\providecommand\@@href[1]{#1\@@endlink}%
\providecommand \@sanitize [0]{\begingroup\catcode`\&12\catcode`\#12\relax}%
\@ifxundefined \pdfoutput {\@firstoftwo}{%
 \@ifnum{\z@=\pdfoutput}{\@firstoftwo}{\@secondoftwo}%
}{%
 \providecommand\@@startlink[1]{\leavevmode}%
 \providecommand\@@endlink[0]{}%
}{%
 \providecommand\@@startlink[1]{%
  \leavevmode
  \pdfstartlink
   attr{/Border[0 0 1 ]/H/I/C[0 1 1]}%
   user{/Subtype/Link/A<</Type/Action/S/URI/URI(#1)>>}%
  \relax
 }%
 \providecommand\@@endlink[0]{\pdfendlink}%
}%
\providecommand \url  [0]{\begingroup\@sanitize \@url }%
\providecommand \@url [1]{\endgroup\@href {#1}{\urlprefix}}%
\providecommand \urlprefix [0]{URL }%
\providecommand \Eprint[0]{\href }%
\@ifxundefined \urlstyle {%
  \providecommand \doi [1]{doi:\discretionary{}{}{}#1}%
}{%
  \providecommand \doi [0]{doi:\discretionary{}{}{}\begingroup
  \urlstyle{rm}\Url }%
}%
\providecommand \doibase [0]{http://dx.doi.org/}%
\providecommand \Doi[1]{\href{\doibase#1}}%
\providecommand \bibAnnote [3]{%
  \BibitemShut{#1}%
  \begin{quotation}\noindent
    \textsc{Key:}\ #2\\\textsc{Annotation:}\ #3%
  \end{quotation}%
}%
\providecommand \bibAnnoteFile [2]{%
  \IfFileExists{#2}{\bibAnnote {#1} {#2} {\input{#2}}}{}%
}%
\providecommand \typeout [0]{\immediate \write \m@ne }%
\providecommand \selectlanguage [0]{\@gobble}%
\providecommand \bibinfo [0]{\@secondoftwo}%
\providecommand \bibfield [0]{\@secondoftwo}%
\providecommand \translation [1]{[#1]}%
\providecommand \BibitemOpen[0]{}%
\providecommand \bibitemStop [0]{}%
\providecommand \bibitemNoStop [0]{.\EOS\space}%
\providecommand \EOS [0]{\spacefactor3000\relax}%
\providecommand \BibitemShut [1]{\csname bibitem#1\endcsname}%
%</preamble>
\bibitem{romanov72}%
  \BibitemOpen
  \bibfield{author}{%
  \bibinfo {author} {\bibfnamefont{G.~N.}\ \bibnamefont{Romanov}}\ and\
  \bibinfo {author} {\bibfnamefont{S.~S.}\ \bibnamefont{Shakhidzhanov}},\ }%
  \bibfield{journal}{%
  \bibinfo {journal} {{JETP Lett.}}\ }%
  \textbf{\bibinfo {volume} {{16}}},\ \bibinfo {pages} {209} (\bibinfo {year}
  {{1972}})%
  \bibAnnoteFile{NoStop}{romanov72}%
\bibitem{kolokolov75}%
  \BibitemOpen
  \bibfield{author}{%
  \bibinfo {author} {\bibfnamefont{A.~A.}\ \bibnamefont{Kolokolov}},\ }%
  \bibfield{journal}{%
  \bibinfo {journal} {JETP Lett.}\ }%
  \textbf{\bibinfo {volume} {21}},\ \bibinfo {pages} {312} (\bibinfo {year}
  {1975})%
  \bibAnnoteFile{NoStop}{kolokolov75}%
\bibitem{callary76}%
  \BibitemOpen
  \bibfield{author}{%
  \bibinfo {author} {\bibfnamefont{P.~R.}\ \bibnamefont{Callary}}\ and\
  \bibinfo {author} {\bibfnamefont{C.~K.}\ \bibnamefont{Carniglia}},\ }%
  \bibfield{journal}{%
  \bibinfo {journal} {J. Opt. Soc. Am.}\ }%
  \textbf{\bibinfo {volume} {{66}}},\ \bibinfo {pages} {775} (\bibinfo {year}
  {{1976}})%
  \bibAnnoteFile{NoStop}{callary76}%
\bibitem{lukosz76}%
  \BibitemOpen
  \bibfield{author}{%
  \bibinfo {author} {\bibfnamefont{W.}~\bibnamefont{Lukosz}}\ and\ \bibinfo
  {author} {\bibfnamefont{P.~P.}\ \bibnamefont{Herrmann}},\ }%
  \bibfield{journal}{%
  \bibinfo {journal} {Opt. Commun.}\ }%
  \textbf{\bibinfo {volume} {{17}}},\ \bibinfo {pages} {192} (\bibinfo {year}
  {{1976}})%
  \bibAnnoteFile{NoStop}{lukosz76}%
\bibitem{cybulski77}%
  \BibitemOpen
  \bibfield{author}{%
  \bibinfo {author} {\bibfnamefont{R.~F.}\ \bibnamefont{Cybulski}}\ and\
  \bibinfo {author} {\bibfnamefont{C.~K.}\ \bibnamefont{Carniglia}},\ }%
  \bibfield{journal}{%
  \bibinfo {journal} {J. Opt. Soc. Am.}\ }%
  \textbf{\bibinfo {volume} {{67}}},\ \bibinfo {pages} {1620} (\bibinfo {year}
  {{1977}})%
  \bibAnnoteFile{NoStop}{cybulski77}%
\bibitem{cybulski83}%
  \BibitemOpen
  \bibfield{author}{%
  \bibinfo {author} {\bibfnamefont{R.~F.}\ \bibnamefont{Cybulski}}\ and\
  \bibinfo {author} {\bibfnamefont{M.~P.}\ \bibnamefont{Silverman}},\ }%
  \bibfield{journal}{%
  \bibinfo {journal} {{J. Opt. Soc. Am.}}\ }%
  \textbf{\bibinfo {volume} {{73}}},\ \bibinfo {pages} {1732} (\bibinfo {year}
  {{1983}})%
  \bibAnnoteFile{NoStop}{cybulski83}%
\bibitem{kolokolov98}%
  \BibitemOpen
  \bibfield{author}{%
  \bibinfo {author} {\bibfnamefont{A.~A.}\ \bibnamefont{Kolokolov}},\ }%
  \bibfield{journal}{%
  \bibinfo {journal} {J. Commun. Technol. Electron.}\ }%
  \textbf{\bibinfo {volume} {43}},\ \bibinfo {pages} {837} (\bibinfo {year}
  {1998})%
  \bibAnnoteFile{NoStop}{kolokolov98}%
\bibitem{kolokolov99}%
  \BibitemOpen
  \bibfield{author}{%
  \bibinfo {author} {\bibfnamefont{A.~A.}\ \bibnamefont{Kolokolov}},\ }%
  \bibfield{journal}{%
  \bibinfo {journal} {Phys. Usp.}\ }%
  \textbf{\bibinfo {volume} {42}},\ \bibinfo {pages} {931} (\bibinfo {year}
  {1999})%
  \bibAnnoteFile{NoStop}{kolokolov99}%
\bibitem{fan03}%
  \BibitemOpen
  \bibfield{author}{%
  \bibinfo {author} {\bibfnamefont{J.}~\bibnamefont{Fan}}, \bibinfo {author}
  {\bibfnamefont{A.}~\bibnamefont{Dogariu}},\ and\ \bibinfo {author}
  {\bibfnamefont{L.~J.}\ \bibnamefont{Wang}},\ }%
  \bibfield{journal}{%
  \bibinfo {journal} {Opt. Express}\ }%
  \textbf{\bibinfo {volume} {11}},\ \bibinfo {pages} {299} (\bibinfo {year}
  {2003})%
  \bibAnnoteFile{NoStop}{fan03}%
\bibitem{willis08}%
  \BibitemOpen
  \bibfield{author}{%
  \bibinfo {author} {\bibfnamefont{K.~J.}\ \bibnamefont{Willis}}, \bibinfo
  {author} {\bibfnamefont{J.~B.}\ \bibnamefont{Schneider}},\ and\ \bibinfo
  {author} {\bibfnamefont{S.~C.}\ \bibnamefont{Hagness}},\ }%
  \bibfield{journal}{%
  \bibinfo {journal} {Opt. Express}\ }%
  \textbf{\bibinfo {volume} {16}},\ \bibinfo {pages} {1903} (\bibinfo {year}
  {2008})%
  \bibAnnoteFile{NoStop}{willis08}%
\bibitem{siegman10}%
  \BibitemOpen
  \bibfield{author}{%
  \bibinfo {author} {\bibfnamefont{A.}~\bibnamefont{Siegman}},\ }%
  \bibfield{journal}{%
  \bibinfo {journal} {Opt. Photon. News}\ }%
  \textbf{\bibinfo {volume} {21}},\ \bibinfo {pages} {38} (\bibinfo {year}
  {2010})%
  \bibAnnoteFile{NoStop}{siegman10}%
\bibitem{koester66}%
  \BibitemOpen
  \bibfield{author}{%
  \bibinfo {author} {\bibfnamefont{C.~J.}\ \bibnamefont{Koester}},\ }%
  \bibfield{journal}{%
  \bibinfo {journal} {IEEE J. Quantum Electron.}\ }%
  \textbf{\bibinfo {volume} {2}},\ \bibinfo {pages} {580} (\bibinfo {year}
  {1966})%
  \bibAnnoteFile{NoStop}{koester66}%
\bibitem{kogan72}%
  \BibitemOpen
  \bibfield{author}{%
  \bibinfo {author} {\bibfnamefont{B.~Y.}\ \bibnamefont{Kogan}}, \bibinfo
  {author} {\bibfnamefont{V.~M.}\ \bibnamefont{Volkov}},\ and\ \bibinfo
  {author} {\bibfnamefont{S.~A.}\ \bibnamefont{Lebedev}},\ }%
  \bibfield{journal}{%
  \bibinfo {journal} {JETP Lett.}\ }%
  \textbf{\bibinfo {volume} {16}},\ \bibinfo {pages} {100} (\bibinfo {year}
  {{1972}})%
  \bibAnnoteFile{NoStop}{kogan72}%
\bibitem{lebedev73}%
  \BibitemOpen
  \bibfield{author}{%
  \bibinfo {author} {\bibfnamefont{S.~A.}\ \bibnamefont{Lebedev}}, \bibinfo
  {author} {\bibfnamefont{V.~M.}\ \bibnamefont{Volkov}},\ and\ \bibinfo
  {author} {\bibfnamefont{B.~Y.}\ \bibnamefont{Kogan}},\ }%
  \bibfield{journal}{%
  \bibinfo {journal} {Opt. Spectrosc.}\ }%
  \textbf{\bibinfo {volume} {35}},\ \bibinfo {pages} {565} (\bibinfo {year}
  {1973})%
  \bibAnnoteFile{NoStop}{lebedev73}%
\bibitem{silverman83}%
  \BibitemOpen
  \bibfield{author}{%
  \bibinfo {author} {\bibfnamefont{M.~P.}\ \bibnamefont{Silverman}}\ and\
  \bibinfo {author} {\bibfnamefont{J.}~\bibnamefont{R.~F.~Cybulski}},\ }%
  \bibfield{journal}{%
  \bibinfo {journal} {J. Opt. Soc. Am.}\ }%
  \textbf{\bibinfo {volume} {73}},\ \bibinfo {pages} {1739} (\bibinfo {year}
  {1983})%
  \bibAnnoteFile{NoStop}{silverman83}%
\bibitem{nistad08}%
  \BibitemOpen
  \bibfield{author}{%
  \bibinfo {author} {\bibfnamefont{B.}~\bibnamefont{Nistad}}\ and\ \bibinfo
  {author} {\bibfnamefont{J.}~\bibnamefont{Skaar}},\ }%
  \bibfield{journal}{%
  \bibinfo {journal} {Phys. Rev. E}\ }%
  \textbf{\bibinfo {volume} {78}},\ \bibinfo {pages} {036603} (\bibinfo {year}
  {2008})%
  \bibAnnoteFile{NoStop}{nistad08}%
\bibitem{skaar06}%
  \BibitemOpen
  \bibfield{author}{%
  \bibinfo {author} {\bibfnamefont{J.}~\bibnamefont{Skaar}},\ }%
  \bibfield{journal}{%
  \bibinfo {journal} {Phys. Rev. E}\ }%
  \textbf{\bibinfo {volume} {73}},\ \bibinfo {eid} {026605} (\bibinfo {year}
  {2006})%
  \bibAnnoteFile{NoStop}{skaar06}%
\bibitem{lakhtakia07}%
  \BibitemOpen
  \bibfield{author}{%
  \bibinfo {author} {\bibfnamefont{A.}~\bibnamefont{Lakhtakia}}, \bibinfo
  {author} {\bibfnamefont{J.~B.}\ \bibnamefont{{Geddes III}}},\ and\ \bibinfo
  {author} {\bibfnamefont{T.~G.}\ \bibnamefont{Mackay}},\ }%
  \bibfield{journal}{%
  \bibinfo {journal} {Opt. Express}\ }%
  \textbf{\bibinfo {volume} {15}},\ \bibinfo {pages} {17709} (\bibinfo {year}
  {2007})%
  \bibAnnoteFile{NoStop}{lakhtakia07}%
\bibitem{ahlfors}%
  \BibitemOpen
  \bibfield{author}{%
  \bibinfo {author} {\bibfnamefont{L.~V.}\ \bibnamefont{Ahlfors}},\ }%
  \emph{\bibinfo {title} {Complex analysis}}\ (\bibinfo {publisher}
  {McGraw-Hill International Editions},\ \bibinfo {year} {1979})%
  \bibAnnoteFile{NoStop}{ahlfors}%
\bibitem{sturrock}%
  \BibitemOpen
  \bibfield{author}{%
  \bibinfo {author} {\bibfnamefont{P.~A.}\ \bibnamefont{Sturrock}},\ }%
  \bibfield{journal}{%
  \bibinfo {journal} {Phys. Rev.}\ }%
  \textbf{\bibinfo {volume} {112}},\ \bibinfo {pages} {1488} (\bibinfo {year}
  {1958})%
  \bibAnnoteFile{NoStop}{sturrock}%
\bibitem{briggs}%
  \BibitemOpen
  \bibfield{author}{%
  \bibinfo {author} {\bibfnamefont{R.~J.}\ \bibnamefont{Briggs}},\ }%
  \emph{\bibinfo {title} {Electron-Stream Interactions with Plasmas}}\
  (\bibinfo {publisher} {MIT Press},\ \bibinfo {year} {1964})%
  \bibAnnoteFile{NoStop}{briggs}%
\end{thebibliography}
%Merlin.mbs v4.21 2009-07-09.
\def\cprime{$'$}
\end{document}